\documentclass[10pt,journal]{IEEEtran}
\usepackage{ifpdf}
\usepackage{cite}
\ifCLASSINFOpdf
  \usepackage[pdftex]{graphicx}
   \graphicspath{{../pdf/}{../jpeg/}}
   \DeclareGraphicsExtensions{.pdf,.jpeg,.png}
\else
\fi
\usepackage{amsmath}
\usepackage{algorithmic}
\usepackage{array}
\newcolumntype{C}[1]{>{\centering\let\newline\\\arraybackslash\hspace{0pt}}m{#1}}
\usepackage{color}
\definecolor{dblue}{rgb}{0,0,0.8}

\usepackage{xcolor}

\usepackage{subfig}
\usepackage{graphicx}
\usepackage{comment}

\usepackage{stfloats}
{}
{}
{}

\usepackage{booktabs, multicol, multirow}
\usepackage{tabularx}
\usepackage{lipsum}

\begin{document}


\title{Multi-Objective PMU Allocation for Resilient Power System Monitoring}
%\title{Resilience Enhancement for Better System Monitoring with Multi-Objective PMU Allocation}

%\title{Resilience Oriented System Monitoring: Multi-Objective PMU Allocation Based on MO-TLBO Algorithm}

%\title{Multi-Objective Resilient PMU Placement: A MO-TLBO Based Approach}

%\title{Preparedness For Resilient Wide Area Monitoring With Multi-Objective PMU Allocation}


\author{\IEEEauthorblockN{Hamed Haggi, Wei Sun, and Junjian Qi}\\
\IEEEauthorblockA{Department of Electrical and Computer Engineering, University of Central Florida, Orlando, FL, USA\\
hamed@ece.ucf.edu, sun@ucf.edu, junjian.qi@ucf.edu}}

\maketitle


\begin{abstract}
Phasor measurement units (PMUs) enable better system monitoring and security enhancement in smart grids. In order to enhance power system resilience against outages and blackouts caused by extreme weather events or man-made attacks, it remains a major challenge to determine the optimal number and location of PMUs. In this paper, a multi-objective resilient PMU placement (MORPP) problem is formulated, and solved by a modified Teaching-Learning-Based Optimization (MO-TLBO) algorithm. Three objectives are considered in the MORPP problem, minimizing the number of PMUs, maximizing the system observability, and minimizing the voltage stability index. The effectiveness of the proposed method is validated through testing on IEEE 14-bus, 30-bus, and 118-bus test systems. The advantage of the MO-TLBO-based MORPP is demonstrated through the comparison with other methods in the literature, in terms of iteration number, optimality and time of convergence.\par
\end{abstract}


% \fi
\IEEEpeerreviewmaketitle

\section{Introduction}
The traditional power grids are in the process of transforming into smart grids with inclusion of intelligent devices. However, the integration of these digitalized devices has driven power networks more complex and vulnerable to cyber-physical-human (CPH) threats \cite{Haggi2019review}. Among these devices, phasor measurement units (PMUs) can provide system operators a set of synchronized phasor measurements for better system monitoring \cite{Sun2018optimum}. As the installation of PMUs in power grids is still an expensive investment, it is not cost effective to install PMUs at each bus in order to achieve accurate monitoring and reach to the full observability \cite{Teimourzadeh2018contingency}.  Therefore, how to allocate PMUs in power systems in order to achieve full system observability in the case of any disruption is still a preeminent challenge.\par

Several research efforts have been dedicated in the optimal PMU placement (OPP) problem, including different optimization problem formulation and solution algorithms. Authors of \cite{Ling2013Hybrid} implement a minimum spinning tree-genetic algorithm (MST-GA) to solve the OPP problem considering the maximum redundancy. In \cite{MAHARI2013optimal}, a binary imperialistic algorithm (BIA) is implemented by considering the impact of zero injection buses (ZIBs) on bulk power systems. Authors in \cite{AHMADI2011optimal} solve the OPP problem with binary particle swarm (BPSO) algorithm by considering the measurement contingencies such as line outage or PMU failure. Authors in \cite{ravindra2020binary} solve the OPP problem with the objective of maximizing the observability via binary harmony search algorithm (BHSA). \par
Moreover, different objectives and indices are considered simultaneously to form the multi-objective PMU placement (MOPP) problem. In \cite{Mazhari2013multi}, a multi-objective OPP problem is investigated based on the cellular learning automata (CLA) algorithm considering the minimum number of PMUs and maximizing the observability. Authors in \cite{Aminifar2016Amulti} propose a multi-objective framework for enhancing the reliability and minimizing the cost of PMU deployment in power systems via NSGA-II algorithm. In \cite{singh2018amulti}, a multi-objective OPP in power systems is studied by implementing the gravitational search algorithm, considering four cases of PMU loss, line outage, ZIB, and normal PMU placement. Furthermore, PMU placement problem is connected with other functions in power system operation and control. For example, a new integer linear programming method is applied to OPP problem in \cite{Sun2018optimum} and \cite{Pal2016apmu} for minimizing the cost of installation for robust static state estimation. A voltage stability constrained optimal simultaneous PMU placement with the goal of enhancing the measurement reliability is presented in \cite{esmaeili2017voltage}. Authors in \cite{Putranto2014Voltage} propose a voltage stability-based PMU placement based on the active power and the angle differences.\par
\vspace{-0.4cm}
Considering the above literature, determination of the appropriate number and location of PMUs has raised the concern of fully system monitoring in the case of any disruption, which is a major challenge from the perspective of power system resilience enhancement (future preparedness). Especially, any PMU loss or line outages could result in voltage instability. This paper aims to formulate and solve the multi-objective PMU placement consisting of minimizing the number of PMUs, maximizing the observability redundancy with the impact of phasing installation, and the voltage stability, in order to provide a more resilient PMU placement (RPP). Additionally, a modified teaching-learning-based optimization (MO-TLBO) algorithm is applied to solve the single and multi-objective PMU placement problem. The advantages of this method is to achieve optimal solutions with less iteration numbers and less computational time compared to existing methods, and its validity is tested on different case studies.\par
\vspace{-0.4cm}
The rest of the paper is organized as follows. The mathematical formulation of multi-objective RPP is presented in Section II, including the quantification of observability and voltage stability index, and different impact factors in MORPP. The solution algorithm of MO-TLBO is introduced in Section III. Section IV presents the simulation results and analysis. Finally, section V concludes the paper.\par
%\vspace{-0.2cm}
\section{Multi-Objective Problem Formulation}\label{sec2}
\subsection{Minimizing the Number of PMUs}
The first objective of MORPP is to determine the minimum number of PMUs with optimal locations to ensure the system observability. The integer linear programming (ILP) formulation is defined as below \cite{Aminifar2010contingency}:
  %\vspace{-0.2cm}
\begin{subequations}\label{ref}
   { \begin{equation}
   \text{min}\; OF_1=\sum_{j=1}^{\text{N}_\text{bus}} x_j \qquad \forall j\subseteq N_{Bus}
   \end{equation}}
  \vspace{-0.3cm}
   { \begin{equation}
   f_i=\sum_{j=1}^{\text{N}_\text{bus}}x_j\;.\; a_{ij} \geq 1 \qquad \forall i,j\subseteq N_{Bus}
   \end{equation}}
\end{subequations}
where $x_j$ is a binary variable, which equals to 1 if bus $j$ is equipped with PMU, otherwise equals to zero. $N_{bus}$ is the set of all network buses. The observability constraint $f_i$ consists of two terms $x_j$ and $a_{ij}$, which $a_{ij}$ is the bus connectivity parameter and can be defined as below:\par
  %\vspace{-0.1cm}
\begin{equation}\label{bus connectivity}
a_{ij}=\left\{\begin{matrix}
1, & \text{if }i\;\text{=\;j} \\
1, & \text{if buses $i$ and $j$ are connected}\\
0, & \text{otherwise}
\end{matrix}\right.
\end{equation}

The aforementioned objective function can be extended to the cost function, by changing the term $x_j$ to $w_j  x_j$. $w_j$ is the installation cost at bus $j$, and defined as $w_j=(1+0.1*n)C$, where $n$ is the number of measurement channels and $C$ is the cost of each PMU\cite{rashidi2015optimal}.\par
 \vspace{-0.3cm}
\subsection{Maximizing the Observability}
One major concern in the PMU allocation problem is to achieve the maximum observability of the entire system even in the case of any disruption. Therefore, the second objective function of MORPP is to maximize the observability. The mathematical formulation is shown as below:
  %\vspace{-0.1cm}
\begin{subequations}
   { \begin{equation}
   \text{max}\; OF_2=\sum_{j=1}^{\text{N}_\text{bus}} f_i \qquad \forall i,j\subseteq N_{Bus}
   \end{equation}}
  \vspace{-0.3cm}
   { \begin{equation}
   f_i=\sum_{j=1}^{\text{N}_\text{bus}}x_j\;.\; a_{ij} \geq 1
   \end{equation}}
     \vspace{-0.5cm}
   { \begin{equation}
   f_i\leq \text{Maximum \;Connectivity \;of \;Bus} \;i 
   \end{equation}}
     \vspace{-0.5cm}
   { \begin{equation}
   \sum_{j=1}^{\text{N}_\text{bus}} x_j \leq TN_{PMU}
      \qquad \forall j\subseteq N_{Bus} 
   \vspace{-0.2cm}
   \end{equation}}
\end{subequations}

where $OF_2$ defines the system observability redundancy, $TN_{PMU}$ is the total number of PMUs based on the availability. Equation (3c) demonstrates the constraint of maximum line connectivity of bus $i$. Additionally, Equation (3d) is to consider that system operators have to install a limited number of PMUs in power system during multiple time stages, due to economic reasons, or communication limitations, which is called phasing installation\cite{dua2008optimal}. In this paper, the impact of phasing installation on observability is also considered.
%\vspace{-0.25cm}

\subsection{Minimizing the Voltage Stability}
\subsubsection{Objective Function}
Voltage stability is one major concern in power system operation. It can be defined as the ability of a power system to keep the voltage of all nodes in the acceptable range under normal condition or in the case of any disruptions. In this paper, the third objective function is defined as minimizing the voltage stability index-based PMU placement, as formulated in the following \cite{arul2016multi}:\par

\vspace{-0.1cm}
   \begin{subequations}\label{ref}
   { \begin{equation}
   \text{min}\; OF_3=\sum_{i\;and\;j=1}^{\text{N}_\text{bus}} VSI_{ij}\;.\;x_j \qquad \forall i,j\subseteq N_{Bus}
   \end{equation}}
   { \begin{equation}
   2\leq VSOI\leq MNB +1
   \end{equation}}
\end{subequations}
where $VSI_{ij}$ is the voltage stability index with the maximum value of 1, and MNB is the maximum number of branches connected to each bus. If the $VSI$ value is near to the unity, it represents that the line is under stressed condition and needs to be relieved. Otherwise, with a small increase in load, which consequently a change in reactive power, the line may be tripped. On the other hand, if its value is small, it shows that the line works properly. The corresponding line is regarded as a critical line. PMUs should be placed at a bus in order to monitor the critical line that has a higher value of $VSI$. Therefore, each of the buses that have higher $VSI$ should be monitored by at least 2 PMUs. Accordingly, the voltage stability observability index (VSOI) can be formulated as equation (4b). The whole system observability index (WSOI) will be defined as the summation values of $VSOI$ and $f_i$. Voltage stability will be considered in PMU placement problems using the index developed as below.\par
\subsubsection{Voltage Stability Index}
In this paper, based on the $\pi$ model of transmission lines that connect two buses with one line, the voltage stability index ($VSI$) can be defined as\cite{golshani2017advanced}:
\begin{equation}
VSI_{mn}=\frac{4\left ( Z_{mn} \right )^2Q_{n} }{ (V_m)^2X_{mn}}\leq 1
\end{equation}
where $Z_{mn}$ and $X_{mn}$ are the line impedance and reactance between buses $m$ and $n$. $V_m$ is voltage at sending bus, and $Q_n$ is reactive power at receiving bus. 
The calculation procedure of $VSI$ follows three steps:\par
\textbf{Step 1:} The $VSI$ of each line is calculated based on the results of optimal power flow using equation (5).\par
\textbf{Step 2:} For any line, increase the reactive power absorption at receiving bus until the $VSI$ value getting close to unity. If the value goes beyond 1, one of the buses connected to the line will experience a voltage drop and cause the system collapse. \par
\textbf{Step 3:} Obtain the maximum value of $VSI$ at each line, and rank the maximum loadability in the ascending order. The smallest maximum-loadability is ranked the highest, which indicates buses should be monitored more than one PMU.\par
\vspace{-0.3cm} 
\subsection{Impact Factors in PMU Allocation Problem}
For enhancing power system monitoring in the case of any disruption, and accordingly improve system resilience, different impact factors are considered in this paper. These factors are impact of ZIBs, contingencies such as line and PMU outage.
\subsubsection{Impact of ZIBs} To consider the impact of zero-injection buses on RPP, an auxiliary binary parameter $s$ is defined, which equals to 1 if it is a zero-injection bus; otherwise 0. In the formulation of the first objective, equation (1b) should be replaced by the following equation:\par
   \vspace{-0.3cm}
\begin{subequations}\label{ref}
   { \begin{equation}
   f_i=\sum_{j=1}^{\text{N}_\text{bus}} x_j\;.\;a_{ij}\;+\;\sum_{j=1}^{\text{N}_\text{bus}}s_j\;.\;a_{ij}\;.\; y_{ij}\geq 1 \qquad \forall j\subseteq N_{Bus}
   \end{equation}}
      \vspace{-0.5cm}
   { \begin{equation}
   s_j=\sum_{j=1}^{\text{N}_\text{bus}}a_{ij}\;.\; y_{ij}
   \end{equation}}
\end{subequations}
where $s_j$ is a parameter related to ZIB, and $y_{ij}$ is an auxiliary variable of buses $i$ and $j$. For each non-ZIB, the number of auxiliary variables is equal to the number of ZIBs connected to that bus. And for each ZIB, the number of auxiliary variables is equal to the number of ZIBs connected to that bus plus one.
\subsubsection{Impact of Measurement Outage}
Power system may experience a contingency, due to equipment failure, natural disasters, etc. One of the major contingencies could be the loss of measurement. To achieve a resilient observable system against PMU losses, each bus must be observed by at least two PMUs to ensure the observability of adjacent lines. In the MOPP formulation, constraint (1b) will be modified as below:\par
 \begin{subequations}
   { \begin{equation}
   f_i\;+\;s_j\geq 2 \qquad \forall j\subseteq N_{Bus}
   \end{equation}}
   \vspace{-0.3cm}
   { \begin{equation}
   s_j=\sum_{j=1}^{\text{N}_\text{bus}}a_{ij}\;.\; y_{ij}
   \end{equation}}
\end{subequations}
\subsubsection{Impact of Line Outage}
Compared to the single PMU loss condition, another common contingency is the single line outage. The impact of single line outage can be added to the MOPP formulation by modifying equations (1b) to (8a) and (8b) as below:\par
\begin{subequations}\label{ref}
   { \begin{equation}
   f_i^k=\sum_{j=1}^{\text{N}_\text{bus}} x_j\;.\;a_{ij}^k\;+\;\sum_{j=1}^{\text{N}_\text{bus}}s_j\;.\;a_{ij}^k\;.\; y_{ij}^k\geq 1 \qquad \forall j\subseteq N_{Bus}
   \end{equation}}
   \vspace{-0.3cm}
   { \begin{equation}
   s_j=\sum_{j=1}^{\text{N}_\text{bus}}a_{ij}^k\;.\; y_{ij}^k
   \end{equation}}
\end{subequations}
where $k$ is the line index. In addition, the binary connectivity parameter when line $k$ is out, $a_{ij}^k$, is defined as $a_{ij}$ if buses $i$ and $j$ are connected with line $k$; otherwise 0.\par

\begin{comment}
\subsection{Impact Factors in PMU Allocation Problem}
For enhancing power system monitoring in the case of any disruption, and accordingly improve system resilience, different impact factors are considered in this paper. These factors are impact of ZIBs, contingencies such as line and PMU outage. The detailed formulation and other information regarding the aforementioned factors can be referred in \cite{Aminifar2010contingency}. In addition, the impact of phasing installation during different time horizons is also studied in this paper.\par
\end{comment}
%\vspace{-0.2cm}
\section{Solution Algorithm}
There are many nature inspired optimization algorithms, such as genetic algorithm (GA), artificial bee colony (ABC), particle swarm optimization (PSO), artificial neural network, etc. These algorithms suffer from issues such as difficulty in determining the optimal controlling parameters like population size, crossover rate and mutation rate, etc. Therefore, a change in the algorithm parameters greatly impacts the effectiveness of the algorithm. However, based on the comparison in \cite{Rao2011Teaching}, the advantage of TLBO algorithm is highly recognized, which obtains global optimal solutions with high consistency and less computational effort, as well as without having any issues mentioned above. In this paper, a modified TLBO algorithm is used for solving single and multi-objective OPP by considering different scenarios.\par
In this study, considering the MO-TLBO algorithm, the students are divided into various groups of learners. Corresponding to the knowledge level of students, different teachers are assigned to these groups, in which teachers try to enhance their own students' knowledge based on their adaptive teaching factors. In the learning phase, the knowledge of students in each group will be updated using the knowledge of other students in the same group as well as by self-learning. Then the major goal of MO-TLBO is to maximize students' knowledge towards teachers' knowledge based on different teaching and self-learning factors \cite{gu2019improved}. \par
%\vspace{-0.15cm}
\section{Simulation Results}
The performance and optimality analysis of the MO-TLBO algorithm on the RPP is validated by testing on different IEEE test cases with various impact factors. Finally, the results for multi-objective PMU placement considering all three objective functions is presented.%All simulations are implemented in a computer with 12 GB RAM, i7 core, and 2.59 GHz CPU.\par
\vspace{-0.3cm}
\subsection{Performance Analysis of Solution Algorithm}
\begin{figure}[b]
\centering
	\includegraphics[width=3.4in,height=1.8in]{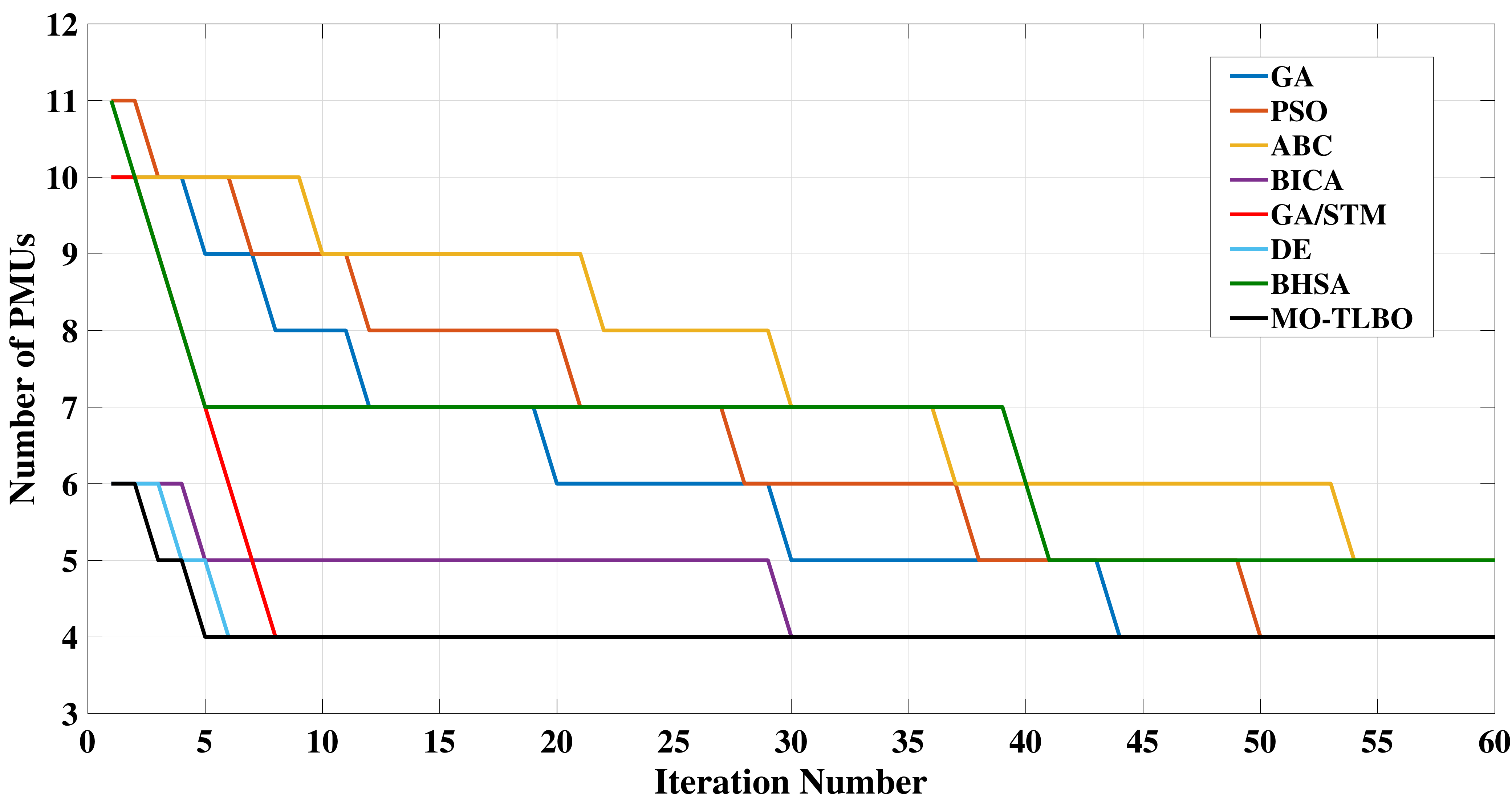}
	\caption{Convergence comparison of different algorithms on IEEE 14-bus test system.}
    \label{Fig. convergence}
\end{figure}

%The MO-TLBO is applied to solve the aforementioned problem. 
In literature, IEEE 14-bus test system without considering impact of ZIBs is a benchmark for testing PMU placement algorithms. The comparison of different solution algorithms, in terms of convergence, is shown in Fig. \ref{Fig. convergence}. It can be seen that MO-TLBO has the advantage of achieving the same optimal solution (e.g. 4 PMUs) in much less iteration number. All other algorithms, including GA, PSO, ABC, hybrid GA and spanning tree method (STM) and binary imperialistic competition algorithm (BICA) need more iterations, while MO-TLBO achieves the convergence after 5 iterations.\par 
%\vspace{-0.1cm}
\par
%\vspace{-0.2cm}

\begin{table}[t]
\caption{Comparison of Total Number of PMUs Using Different Algorithms Considering ZIBs}
{\begin{tabular*}{21pc}{@{\extracolsep{\fill}}lccc@{}}\toprule \label{comparison}
Algorithm  &IEEE 14-bus &IEEE 30-bus & IEEE 118-bus \\
%heading  &heading two &  three \\
\hline
%GA \cite{mohammadi2009optimal}  &3 &7  &29 \\
ILP  \cite{Aminifar2010contingency}   &3 &-  &29 \\
ABC  \cite{li2017optimal}   &3 &8  &- \\
PSO  \cite{AHMADI2011optimal}   &3 &7  &29 \\
BHSA  \cite{ravindra2020binary}   &3 &8  &- \\
BICA  \cite{MAHARI2013optimal}   &3 &7  &28 \\
IGA  \cite{aminifar2009optimal}   &3 &7  &29 \\
GA/STM  \cite{devi2020hybrid}  &3  &-  &29\\
MO-TLBO  &3 &7  &28 \\
\hline
\end{tabular*}}{}
\end{table}
\vspace{-0.2cm}
\subsection{Optimality Analysis through Comparisons}
In order to analyze the optimality of MO-TLBO algorithm, Table \ref{comparison} summarizes the results of minimizing the number of PMUs considering the ZIBs, using different methods. Additionally, the best locations to equip PMUs considering ZIBs are also shown in Table \ref{ZIB}. Total nine different solution algorithms are tested against IEEE 14-, 30- and 118-bus systems. It is shown that the MO-TLBO algorithm achieves the optimal solutions with the fastest convergence speed.\par
\begin{table}[t]
\caption{Results of Minimizing PMUs Numbers With ZIBs}
{\begin{tabular*}{21pc}{@{\extracolsep{\fill}}lcc@{}}\toprule \label{ZIB}
Test system  & Considering ZIBs \\
%heading  &heading two &  three \\
\midrule
IEEE \;14 \;Bus   &2,6,9  \\
\\
IEEE \;30  \;Bus   &1,7,10,12,18,23,27  \\
\\
IEEE 118 Bus   &2,9,11,13,17,21,25,28,34,37,40,45,49,53,56,62,72,75,77\\
&80,85,86,90,94,102,105,110,114\\
\hline
\end{tabular*}}{}
\end{table}
%\vspace{-0.3cm}
\subsection{Results of the Impact of Measurement Loss}
In the case of any PMU loss, all other lines should be monitored by at least one secondary PMU in the neighborhood. This redundancy requirement will lead to locate as many PMUs as possible to address system resilience from the monitoring perspective. Table \ref{Loss} presents the results of minimizing the number of PMUs in the case of any measurement loss. Compared to the base case with ZIBs, more PMUs are installed to monitor the system and maintain the system observable in the case of any measurement loss, including 4, 5, and 33 more PMUs for IEEE 14-, 30- and 118-bus systems, respectively.\par 
%\vspace{-0.3cm}
\begin{table}[t]
\caption{Results of Impact of Measurement Loss}
{\begin{tabular*}{21pc}{@{\extracolsep{\fill}}lcc@{}}\toprule \label{Loss}
Test System  &The Impact of Measurement Loss\\
\hline
IEEE \;14 \;Bus  &1,4,5,6,9,10,13  \\
\\
IEEE \;30  \;Bus  &1,3,4,7,10,12,15,16,18,19,20,24,27,29,30  \\
\\
IEEE 118 Bus &1,3,6,8,9,11,12,15,17,19,20,21,23,26,27,28,29,32,34,35,40\\

&42,44,45,46,49,51,52,54,56,57,59,62,66,68,70,71,75,76,77\\

&78,80,83,85,86,87,89,91,92,94,96,100,101,105,106,108,110\\

&111,112,115,117 \\

\hline
\end{tabular*}}{}
\end{table}
\vspace{-0.25cm}
\subsection{Results of the Impact of Line Outage}
Table \ref{line} summarizes the results of minimizing the number of PMUs in the case of line outage. Taking IEEE 30-bus test system as an example, 13 PMUs are needed to observe the system with line outage. Compared to the base case, 6 more PMUs are added to the system to maintain the observability. Moreover, as the impact of line outage is less severe than the impact of measurement loss, it requires less number of extra PMUs for the line outage, as shown in Table \ref{Loss} and Table \ref{line}.\par
\vspace{-0.1cm}
\begin{table}[t]
\caption{Results of Impact of Line Outage}
{\begin{tabular*}{21pc}{@{\extracolsep{\fill}}lccc@{}}\toprule \label{line}
Test System  &The Impact of Line Outage\\
\midrule
IEEE \;14 \;Bus  &1,3,6,10,11,13  \\
\\
IEEE \;30  \;Bus  &1,3,4,10,12,13,15,16,17,19,23,26,29 \\
\\
IEEE 118 Bus &1,6,10,11,12,15,17,19,21,23,24,25,27,29,32,34,35,40,42,44\\

&46,49,51,53,56,5,59,62,63,70,73,75,76,78,80,83,85,87,89\\

&91,92,94,96,100,102,105,106,109,111,112,115,166,117\\

\hline
\end{tabular*}}{}
\end{table}
\subsection{Results of the Impact of Phasing on Observability}
\begin{figure}
\centering
	\includegraphics[width=3.5in,height=1.8in]{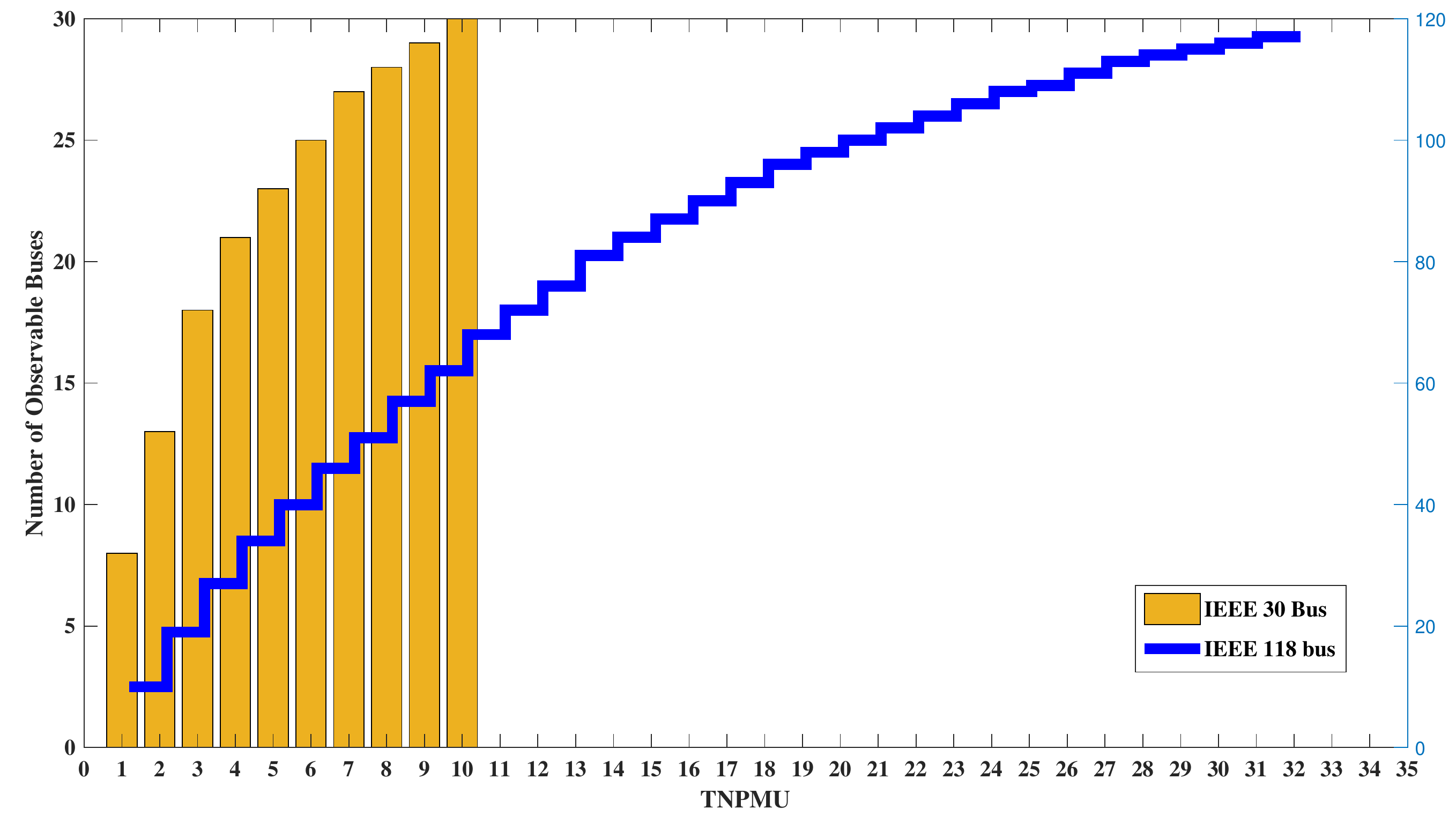}
	\caption{The effect of Phasing in OPP for different case studies.}
    \label{Fig. The effect of Phasing in OPP}
\end{figure}

Considering the second objective function, the results are presented in Table \ref{Phasing} and Fig. \ref{Fig. The effect of Phasing in OPP}. Take IEEE 118-bus system as an example, as shown in Fig. \ref{Fig. The effect of Phasing in OPP}, it is assumed that the network planner has limited budget to install 32 PMUs in the network through 3 time stages. At the first stage, 11 PMUs are installed to enable total 72 buses observable, which achieves approximately 61\% observability of the system. Next in the second stage, with additional 13 PMUs installed, total 110 buses can be observable to achieve about 91\% observability of the system. Finally in the last stage, with another 8 PMUs installed, all 118 buses are observable with 100\% observability of the system. The phasing installation enables system planners to install PMUs in critical areas required to be monitored. Consequently, it assists planners to evaluate the monetary predictions of installing PMUs to reach maximum observability at each time stage.\par
\begin{table}
\caption{Impact of Phasing in IEEE 14-Bus System}
{\begin{tabular*}{21pc}{@{\extracolsep{\fill}}lccc@{}}\toprule \label{Phasing}
$TN_{PMU}$  &Bus Location  & Observability (\%) \\
%heading  &heading two &  three \\
\midrule
1  &6 &0.4286\%  \\
2  &6,9 &0.7143\%   \\
3  &2,6,9 &0.9286\%  \\
4  &2,6,7,9 &100\%  \\

\hline
\end{tabular*}}{}
\end{table}
\vspace{-0.3cm}
\subsection{Results of System Redundancy}
The redundancy values of system observability for each of three IEEE test cases are presented in Table \ref{redundant}. Considering the benchmark, the system redundancy value is lower than the second scenario considering ZIBs. The reason is ZIBs improve the system observability, but prevent from any additional redundant measures from PMUs. For the scenario of line outage and PMU loss, the system observability redundancy value is greater than first two scenarios. The reason is that in the case of any disruption, each of the PMUs should take the responsibility of observing the neighboring lines. As a result, each PMU measures more than one time, and consequently the system observability redundancy increases.\par
\begin{table}
\caption{System Redundancy in Different Scenarios}
{\begin{tabular*}{21pc}{@{\extracolsep{\fill}}lcccc@{}}\toprule \label{redundant}
Test System  &Ignoring  & Considering &Line &PMU \\
          &ZIB  & ZIB &Outage &Loss \\
%heading  &heading two &  three \\
\midrule
IEEE 14 Bus  &19 &16 &34  &34 \\
IEEE 30 Bus &50 &42  &56  &59 \\
IEEE 118 Bus  &178 &159  &245 &288\\

\hline
\end{tabular*}}{}
\end{table}
%\vspace{-0.1cm}
\subsection{Results of Multi-Objective PMU Placement}
Due to the page limit, the results of voltage stability analysis is only presented for IEEE 14-bus system in Table \ref{VSI14}. The procedure of calculating $VSI$ is same for other two test systems as discussed in section II, part C. The tables are arranged in the ascending order of maximum reactive power loadability. The results of MOPP considering all objective functions are presented in Table \ref{OF123}. Taking the IEEE 14-bus system as an example, the most critical buses are the ones with the minimum limit of the load increase. In order to monitor these buses properly in every possible situation, they must be observed by at least 2 PMUs. This means the observability redundancy should be at least 2 for these weak buses and demonstrates the fact that these buses can be fully observable, even in the case of any disruption. \par
\begin{table}
\caption{Results of VSI Calculation - IEEE 14-Bus System}
{\begin{tabular*}{21pc}{@{\extracolsep{\fill}}lccc|lccc@{}}\toprule \label{VSI14}
Load  &Qmax  &VSI  &Critical &Load  &Qmax  &VSI  &Critical\\
Bus   &(pu)     &Index     &Line  &Bus   &(pu)     &Index     &Line\\
%heading  &heading two &  three &four \\
\midrule
9  &0.436 &0.9874 &4-9 &6 &0.985 &0.9902 &5-6\\

13 &0.538 &0.9966 &12-13 &4  &1.159 &0.9924 &3-4\\

14 &0.540 &0.9941 &13-14 &5  &1.186 &0.9970 &1-5\\

12 &0.826 &0.9998 &6-12 &3  &1.280 &0.9795 &2-3\\

11 &0.908 &0.9903 &10-11 &10 &1.528 &0.9892 &9-10\\

\hline
\end{tabular*}}{}
\end{table}
\vspace{-0.2cm}
\begin{table}[t]
\caption{Results of Multi-objective PMU Placement}
{\begin{tabular*}{21pc}{@{\extracolsep{\fill}}lcccc@{}}\toprule \label{OF123}
Test System  &Multi-Objective  &WSOI \\
Test System  &PMU Locations & \\
\midrule
IEEE \;14 \;Bus  &2,6,9,12,14 &23\\
\\
IEEE \;30  \;Bus  &3,4,10,12,14,15,18,24,26,29,30  &76\\
\\
IEEE 118 Bus &1,2,9,12,13,18,21,26,28,29,34,39,42,45,47 &279\\

&49,51,54,59,62,69,70,72,77,85,87,90,91,94\\

&99,100,101,106,110,114\\
\hline
\end{tabular*}}{}
\end{table}
%\vspace{-0.3cm}
\section{Conclusion}
In this paper, a new modified algorithm of MO-TLBO is implemented on single and multi-objective PMU placement problems with various objective functions. The algorithm is tested on three IEEE test systems and compared with other best algorithms in the literature. In order to achieve a resilient PMU placement, a multi-objective RPP is solved based on minimizing VSI and the number of PMUs, and maximizing observability. Simulation results demonstrate that the proposed MO-TLBO algorithm present better results with less iteration number and less computational time.\par
\vspace{-0.1cm}

\bibliographystyle{IEEEtran}
\bibliography{main}

% thats all folks
\end{document}